\documentstyle[aps,pra,psfig,epsf,epsfig]{revtex}
\begin{document}
\renewcommand {\thefootnote} {\fnsymbol{footnote}}
\twocolumn
\draft
\title{Least-squares inversion for density-matrix reconstruction
}
\author{
T. Opatrn\'{y}
\cite{Olomouc}
and D.-G. Welsch
 }
\address{
Friedrich-Schiller-Universit\"{a}t Jena,
Theoretisch-Physikalisches Institut
\\
Max-Wien-Platz 1, D-07743 Jena, Germany 
}
\author{
W. Vogel
 }
\address{
Universit\"{a}t Rostock, Fachbereich Physik, Arbeitsgruppe Quantenoptik
\\
Universit\"{a}tsplatz 3, D-18051 Rostock, Germany 
}
\date{March 10, 1997}
\maketitle
\begin{abstract}
We propose a method for reconstruction of the density matrix 
from measurable time-dependent (probability) distributions of 
physical quantities. 
The applicability of the method based on least-squares inversion
is -- compared with other methods -- very universal. It can be used 
to reconstruct quantum states of 
various systems, such as harmonic and and anharmonic oscillators 
including molecular vibrations in vibronic transitions and damped motion. 
It also enables one to take into account various specific features of 
experiments, such as limited sets of data and data smearing owing to 
limited resolution. 
To illustrate the method, we consider a Morse oscillator and 
give a comparison with 
other state-reconstruction methods suggested recently.
\end{abstract}


%%%%%%%%%%%%%%%%%%%%%%%%%%%%%%%%%%%%%%%%%%%%%%%%%%%%%%%%%%%%%%%%%%%%%

\section{Introduction}
\label{S1}

During the last years the problem of quantum-state measurement has been of
increasing interest. Advances in the experimental techniques,
which have allowed the physicists to measure not only single
observables of a quantum-mechanical system but -- for certain systems --
the quantum state as a whole, have stimulated both experimental
and theoretical research. A number of proposals for measuring quantum
states have been made and various quantum-mechanical systems have
been considered.

In quantum optics balanced homodyning has been a very
fruitful method for tomographical reconstruction \cite{Vogel Risken} of the
quantum state of optical fields. Measuring the quadrature-component
statistics and using inverse Radon transform techniques,
the method was successfully applied to the determination
of the Wigner function of single-mode signal fields \cite{Wigner}.
The method (also called optical homodyne tomography) has also been extended
to direct sampling of the density matrix in a quadrature-component basis
\cite{Kuehn} and in the Fock basis \cite{Ariano,Schiller}.
Tomographical methods can also be used for reconstructing
the quantum state of matter systems, such as molecular vibrations
\cite{Dunn2} or the transverse motion of an atom beam \cite{Janicke}. 
In the case of molecular vibrations the time- and frequency-resolved
fluorescence spectrum of the molecule plays the role of the
quadrature components, provided that the vibrational frequencies
in the two electronic states are almost equal to each other
and the vibrational motion is approximately harmonic.
Further, tomographical methods have been considered in the reconstruction
of the (harmonic) center-of-mass motion of trapped atoms
\cite{Wa-Vo1}.

Including in the balanced detection scheme additional vacuum
inputs, the quantum state of optical fields can be directly measured
in terms of phase-space functions \cite{Walker}. Using unbalanced
homodyning, the displaced photon-number statistics of the
signal field can be measured, from which the quantum state can
be obtained in terms of phase-space functions \cite{Wa-Vo2}
and the density matrix in the Fock basis \cite{Tom}.
The method also applies to matter systems and was used to
reconstruct the quantum state of the (harmonic) center-of mass motion
of trapped ions \cite{Leibfried}. In this case the coherent
displacements are induced by rf fields that are resonant to
the motional frequency.

The methods considered so far are restricted to undamped harmonic
oscillators. However, in many physical systems anharmonic motions
(modified by dampings) are observed.
Recently, interesting nonclassical quantum states have been prepared in
systems that are typically anharmonic. Examples are the generation of
amplitude-squeezed states in molecules~\cite{Dunn2,Dunn1} and of
Schr\"odinger cat-like states in atomic Rydberg wave packets~\cite{Noel}.
A first attempt has been made to reconstruct the quantum state of
anharmonic molecular vibrations using time-resolved fluorescence
spectroscopy \cite{Shapiro}. It has been shown that the density
matrix can be obtained by inversion of high-dimensional systems of
linear equations. An approach has been given in
\cite{LeoRaymer}, extending the pattern-function formalism to
more general than harmonic potentials and reconstructing the density matrix
from the time-dependent position distribution.
However, the quantities that are typically measured in molecular
spectroscopy are different from position distributions in general.
For example, the time-gated fluorescence spectrum measured
in experiments of the type described in \cite{Dunn2}
is determined by the Franck-Condon overlap factors of the vibrational
wave functions in two electronic states. For anharmonic vibrations 
this spectrum cannot be identified with distributions of the type of
position distributions and therefore it should be used directly 
in order to reconstruct the density matrix \cite{Kowalczyk}.

In any case there have been a number of open questions, such
as those of the determination of suitable sampling functions mapping the
measured data onto the density matrix, the choice of optimum observational
times, and the inclusion into the scheme of damping effects and data
smearing. The aim of the present paper is to offer possible answers to
some of them. For all the systems mentioned
the general problem to be solved is the inversion of linear equations
that relate the measured quantities to the density-matrix elements
of the system under study. This can be done in a very effective
way using the least-squares method, which dates back to the eras of
Legendre and Gauss \cite{Gauss}. Actually, the method has been used in
the context of quantum-state reconstruction problems, such as the
reconstruction of the quantum state of cavity fields by quantum-state
endoscopy \cite{Schleich}, vibrations of trapped ions \cite{Leibfried},
and optical field by balanced \cite{Tan} and unbalanced \cite{Tom}
homodyning.

For comparison with previous work \cite{LeoRaymer,Leo-mors}, we will restrict
attention to the reconstruction of the density matrix from the
time-dependent position distribution of a particle moving in an
anharmonic potential. We show that the least-squares method can
advantageously be used in order to reduce the statistical error. We
further show that the flexibility of the method enables us to perform the
reconstruction from data recorded during shorter time intervals and to
take into account typical experimental features, such as smeared and
imperfect data. Finally, the method can also be used to reconstruct
quantum states of damped systems. It is worth noting that
the applicability of the method only requires that there
are measurable quantities that are linear combinations of all
the relevant density-matrix elements of the quantum state of the
system under study.

The paper is organized as follows. In Sec.~\ref{S2} the problem
of construction of sampling functions is considered. Sections \ref{S3}
and \ref{S4}, respectively, are devoted to the questions of observational
time and data smearing. Concluding remarks are given in Sec.~\ref{S5}.
Elements of the least-squares method and the statistical-error analysis
are outlined in Apps.~\ref{Ap1} and \ref{Ap2}.

%%%%%%%%%%%%%%%%%%%%%%%%%%%%%%%%%%%%%%%%%%%%%%%%%%%%%%%%%%%%%%%%%%%

\section{Sampling functions}
\label{S2}

Let us consider a quantum-mechanical system and assume
that at some initial time it is prepared in a state with
density matrix $\varrho_{n,n'}$ $\!=$ $\!\langle n | \hat{\varrho}
| n' \rangle$, where $| n \rangle$ are the energy eigenstates of the 
system Hamiltonian. Further, let us assume that there is a measurable
time-dependent (probability) distribution $p(x,t)$ of a quantity $x$
that can be given by a
linear combination of all density-matrix elements $\varrho_{n,n'}$
that are initially excited, with linearly independent coefficient
functions $S_{n,n'}(x,t)$,
\begin{eqnarray}
\label{t0}
p(x,t) = \sum_{n,n'} S_{n,n'}(x,t) \, \varrho _{n,n'}.
\end{eqnarray}
When we consider, e.g., a particle that moves in a potential well
and is initially prepared in a bound state
(e.g., a molecular vibration in a vibronic system below
the dissociation level), only the discrete part of
the energy spectrum is excited ($n$ $\!=$ $\!1,2,3,\ldots$).
For the sake of transparency, in what follows we restrict attention to
discrete spectra. However, replacing in Eq.~(\ref{t0}) the sums with
integrals (or combinations of sums and integrals), excitations
of continuous parts of the spectrum can be treated accordingly.
For any physical state the density-matrix elements $\varrho_{n,n'}$
must eventually decrease indefinitely with increasing $n(n')$. Therefore it
follows that the expression on the right-hand side of Eq.~(\ref{t0})
can always be approximated to any desired degree of accuracy by setting
\begin{eqnarray}
\label{t0a}
\varrho _{n,n'} \approx 0 \quad {\rm for} \quad n(n') > n_{\rm max},
\end{eqnarray}
if $n_{\rm max}$ is suitably large. From the assumptions made it is
is clear, that Eq.~(\ref{t0}) can, in principle, be inverted
in order to obtain the quantum state from the measured function $p(x,t)$.

A powerful method for solving the problem is least-squares inversion.
To illustrate the method, let us us suppose that $p(x,t)$ is a
(one-dimensional) position probability of a particle moving
in a potential well. When during the
time interval of observation the temporal evolution of the quantum state
of the particle is only governed by the system Hamiltonian, then
Eq.~(\ref{t0}) together with Eq.~(\ref{t0a}) can be written as
\begin{eqnarray}
\label{t1}
p(x,t) = \sum_{n,n' \le n_{\rm max}}
S_{n,n'}(x,t)
%e^{ i (\omega_{n} - \omega_{n'}) t } \psi_{n}(x) \psi_{n'}(x)
\, \varrho _{n,n'} ,
\end{eqnarray}
where $S_{n,n'}(x,t)$ is given by
\begin{eqnarray}
\label{t1a}
S_{n,n'}(x,t)
=  \psi_{n}(x) \psi_{n'}(x) \, e^{- i (\omega_{n} - \omega_{n'}) t }.
\end{eqnarray}
Here $\omega_{n}\!-\!\omega_{n'}$ are the transition frequencies
of the system, the energy eigenfunctions in the 
position representation being given by $\psi_{n}(x)$.

%%%%%%%%%%%%%%%%%%%%%%%%%%%%%%%%%%%%%%%%%%%%%%%%%%%%%%%%%%%%%%%%%%%%%%%

\subsection{Harmonic oscillators}

The simplest system is an undamped harmonic oscillator of frequency 
$\omega_{\rm h}$. The equidistant energy spectrum, 
$\omega_{n}\!-\!\omega_{n'}$ $\!=$ $(n\!-\!n')\omega_{\rm h}$,
enables us to separate single diagonals of the density matrix.
Introducing the Fourier transform
\begin{eqnarray}
\label{t2}
p^{(k)}(x) = \frac{1}{T} \int_{0}^{T} d{t} \,
e^{ i \omega_{k} t} p(x,t),
\end{eqnarray}
where $\omega_{k}$ $\!=$ $\!k\omega_{\rm h}$, $k$ $\!=$ $\!0,1,2,\ldots$, 
and $T$ $\!=$ $\!2\pi$, we can express $p^{(k)}(x)$ in terms of
$\varrho_{n,n+k}$ as
\begin{eqnarray}
\label{t3} 
p^{(k)}(x) = \sum_{n=0}^{n_{\rm max}-k} 
 \psi_{n+k}(x) \psi_{n}(x) \, \varrho_{n+k,n}
\end{eqnarray}
[$\psi_{n}(x)$ $\!=$ $\!(\sqrt{\pi} 2^{n} n!)^{-1/2}$
$\!\exp (-x^{2}/2) {\rm H}_{n}(x)$, $x$ being dimensionless].
Using the least-squares method, Eq.~(\ref{t3}) can easily be
inverted to obtain the density-matrix elements in terms of the
measured quantities. Comparing Eq.~(\ref{t3}) with Eq.~(\ref{e1})
and applying Eq.~(\ref{e6a}), the reconstructed density
matrix elements $\tilde \varrho_{n+k,n}$ are given by
\begin{eqnarray}
\label{t4}
\tilde \varrho_{n+k,n} = \int {d}x \, K_{n}^{(k)}(x) \, p^{(k)}(x) ,
\end{eqnarray}
where
\begin{eqnarray}
\label{t5}
K_{n}^{(k)}(x) = \sum_{l=0}^{n_{\rm max}-k}
F^{(k)}_{n,l} B_{l}^{(k)}(x).
\end{eqnarray}
Here, the matrix ${\bf F}^{(k)}$ is the inverse of the matrix
${\bf G}^{(k)}$, ${\bf F}^{(k)}$ $\!=$ $\!({\bf G}^{(k)})^{-1}$,
with the matrix ${\bf G}^{(k)}$ being defined by
\begin{eqnarray}
\label{t7}
G^{(k)}_{m,n} = \int {d}x \, B_{m}^{(k)}(x) B_{n}^{(k)} (x)
\end{eqnarray}
($m,n = 0,1,\dots n_{\rm max}\!-\!k$), and
\begin{eqnarray}
\label{t6}
B_{l}^{(k)}(x) = \psi_{l+k}(x) \psi_{l}(x).
\end{eqnarray}
  From Eqs.~(\ref{t5}) -- (\ref{t6}) it follows that
\begin{eqnarray}
\label{t7aa}
\int {d}x \, K^{(k)}_{n}(x) B^{(k)}_{n'} (x) = \delta_{n,n'} .
\end{eqnarray}

It was found \cite{Tan} that when $n$ $\!\ll$ $\!n_{\rm max}$
then the integral kernels (\ref{t5}) are essentially identical
to the sampling (pattern) functions derived in \cite{Ariano}.
Moreover, it can be shown that only for
values of $n(n')$ that are close to $n_{\rm max}$ the two methods
yield substantially different sampling functions, the oscillating
behavior of the least-squares sampling functions being less pronounced
than those in \cite{Ariano}. This suggests that the statistical
error of the reconstructed density-matrix elements $\tilde \varrho_{n,n'}$
with $n(n')$ close to $n_{\rm max}$ is smaller for the
least-squares method than for the method in \cite{Ariano},
because the statistical fluctuation depends on the squares of the
sampling functions [see Eq.~(\ref{e14})]. The decrease of the statistical
error can be understood as a consequence of the {\em a priori}
information on the state to be reconstructed: 
the reconstructed elements $\tilde \varrho_{n,n'}$ with $n(n')$ close to
$n_{\rm max}$ cannot be ``contaminated'' by neighboring elements with
indices above $n_{\rm max}$.
Clearly, when the probability of finding excited levels above $n_{\rm max}$ 
is not negligibly small, then the least-squares method can cause a 
systematical error. Taking into account the complete set of density-matrix 
elements, we have, in place of Eq.~(\ref{t3}),
\begin{eqnarray}
\label{t7a}
\lefteqn{
p^{(k)}(x) = \sum_{n=0}^{\infty}  \psi_{n+k}(x) \psi _{n}(x)
\, \varrho _{n+k,n}
}
\nonumber \\  && \quad
= \left( \sum_{n=0}^{n_{\rm max} - k} +
\sum_{n=n_{\rm max} - k}^{\infty}   \right)
  \psi_{n+k}(x) \psi _{n}(x) \, \varrho _{n+k,n}.
\end{eqnarray}
Using Eqs.~(\ref{t4}) and (\ref{t7aa}), we derive
\begin{eqnarray}
\label{t7b}
\lefteqn{
\tilde \varrho_{n,n+k} = \varrho_{n,n+k}
}
\nonumber \\ && \quad
+ \!\!\sum_{n'=n_{\rm max}-k+1}^{\infty} \varrho _{n'+k,n'}
\int {d}x \, K^{(k)}_{n}(x) \psi_{n'+k}(x) \psi_{n'}(x),
\nonumber \\ &&
\end{eqnarray}
where the second term represents the systematical error.

%%%%%%%%%%%%%%%%%%%%%%%%%%%%%%%%%%%%%%%%%%%%%%%%%%%%%%%%%%%%%%%%%%%%

\subsection{Anharmonic systems}

In order to treat the more general case of non-equidistant energy
levels, Eq.~(\ref{t3}) can be generalized to
\begin{eqnarray}
\label{t3a}
p^{(k)}(x)
= \sum_{n,n'\leq n_{\rm max} \atop \omega_{n}-\omega_{n'}=\omega_{k}}
\psi_{n}(x) \psi_{n'}(x) \, \varrho_{n,n'},
\end{eqnarray}
where $p^{(k)}(x)$ is defined by
\begin{eqnarray}
\label{t3ab}
p^{(k)}(x) = \lim_{T \to \infty} \frac{1}{T}
\int_{0}^{T} dt \ e^{i \omega _{k} t} p(x,t) .
\end{eqnarray}
We see that it is impossible in general to separate single diagonals
of the density matrix by integrating the position probability over some
finite time interval. Actually, integration over an infinite time interval
needs doing in order to exactly single out terms oscillating
at chosen transition-frequencies of the system, which is of course
illusory. We will study this problem in Sec.~\ref{S3} in more detail.
Here we assume that the terms are (approximately) singled out and the
density-matrix reconstruction can be performed inverting Eq.~(\ref{t3a})
in the same way as for the harmonic oscillator, i.e.,
\begin{eqnarray}
\label{tad1}
\tilde \varrho_{n,n'} = \int {d}x \, K_{n,n'}(x) \, p^{(k)}(x) ,
\end{eqnarray}
where
\begin{eqnarray}
\label{tad3}
K_{n,n'}(x) = 
\sum_{m,m'\leq n_{\rm max} \atop \omega_{m}-\omega_{m'}=\omega_{k}}
F_{n,n';m,m'} B_{m,m'} (x) .
\end{eqnarray}
The double-indices $n,n'$ $(m,m')$ are
to be chosen such that $\omega_{n(m)}$ $\! -$ $\! \omega_{n'(m')}$
$\! =$  $\!\omega_{k}$, and $n,n'$ $\!(m,m')$ $\! \le$ $\! n_{\rm max}$.
For chosen $k$ the matrix ${\bf F}$ is given by ${\bf F}$ $\! = $
$\!{\bf G}^{-1}$, where
\begin{eqnarray}
\label{tad4}
G_{n,n';m,m'} = \int dx \ B_{n,n'}(x) B_{m,m'} (x),
\end{eqnarray}
\begin{eqnarray}
\label{tad5}
B_{n,n'}(x) = \psi_{n}(x) \psi_{m}(x) .
\end{eqnarray}

\begin{figure}[htb]
\epsfysize=23cm
\epsfbox{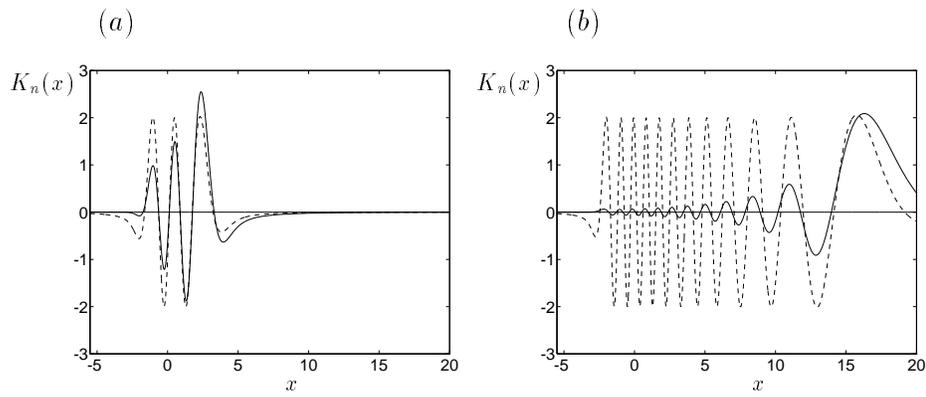}
\vspace{-2cm}
\caption{
Sampling functions for reconstructing the diagonal density-matrix
elements $\varrho_{n,n}$ of a Morse oscillator ($a$ $\!=$ $\! 0.279$)
from the position distribution for $n$ $\!=$ $\!2$ (a) and $n=11$ (b);
full line: least-squares method; dashed line: IWM.
\label{F2}
}
\end{figure}

To give an example, let us consider the bound motion of a particle
in an anharmonic potential of the Morse type,
\begin{eqnarray}
\label{t8}
U(x) = \frac{1}{2a^{2}} \left( e^{-ax} - 1 \right) ^{2}
\end{eqnarray}
($a$ $\!>$ $\!0$). There are $n_{M}$ bound states, where
$n_{M}$ $\! = $ $\![[ a^{-2} - 1/2]]$, with $[[ y ]]$ being the integral
part of $y$.
Their wave functions 
are given by $\psi_{n}(x)$ $\!=$ $\! N_{n} e^{-z/2} z^{b/2} L_{n}^{b}(z)$
($n$ $\!=$ $\!0,1,\dots n_{\rm M}$), where $z$ $\!=$ $\!2a^{-2}e^{-ax}$,
$b$ $\!=$ $\!2a^{-2}-2n-1$, $N_{n}^{2}$ $\!=$ $\! abn!/\Gamma (n+b+1)$
\cite{Morse}. Restricting attention to bound states we can work with
$n_{\rm max}$ $\!\leq$ $\!n_{M}$. 
In Fig.~\ref{F2} we have plotted examples of sampling functions $K_{n}(x)$
$\!\equiv$ $\!K_{n,n}(x)$
that are required for the determination of the diagonal density-matrix
elements $\varrho_{nn}$ [i.e., $\omega_{k}$ $\!=$ $\!0$ in
Eqs.~(\ref{t3a})
and (\ref{t3ab})], including into the calculation all bound states (i.e.,
$n_{\rm max}$ $\!=$ $\!n_{M}$), which exactly corresponds
to the case considered in \cite{Leo-mors} applying the 
irregular wave-function method (IWM) given in \cite{LeoRaymer}. Comparing
the sampling functions with those of IWM, from Fig.~\ref{F2}
we see that the latter yields sampling functions that show
substantially larger oscillations than those obtained by means of
least-squares inversion. In particular,
we see that the effect is much stronger than for harmonic oscillators.
A reconstruction based on least-squares inversion is therefore
expected to give rise to substantially smaller statistical fluctuations
than IWM.

\begin{figure}[htb]
\epsfysize=23cm
\epsfbox{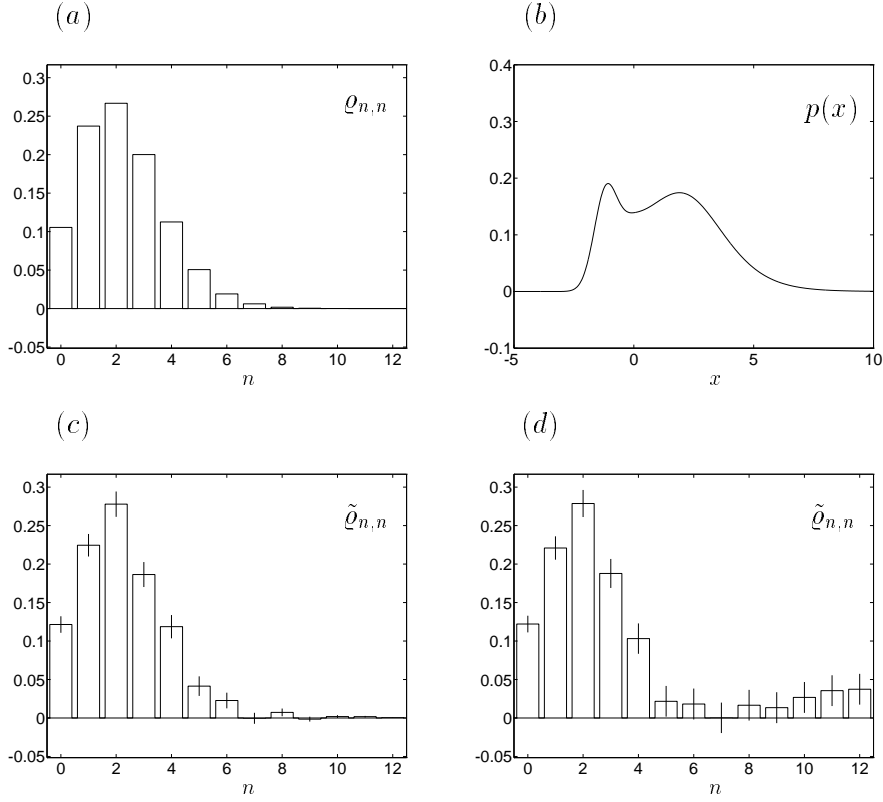}
\vspace{-2cm}
\caption{
Reconstruction of the diagonal density-matrix elements
of a Morse oscillator ($a$ $\!=$ $\! 0.279$), which is assumed to be
prepared in a state $\langle n | \psi \rangle$  $\! \propto$
$\!\alpha^{n}n!^{-1/2}$, $\alpha$ $\! =$ $\!-1.5$, from
$N_{\rm e}$ $\!=$ $\!5\times 10^{3}$ recorded events
in a simulated position measurement;
exact density-matrix elements $\varrho_{n,n}$ (a),
exact time-averaged position distribution (b),
density-matrix elements $\tilde\varrho_{n,n}$ reconstructed using
least-squares inversion (c) and IWM (d).
The error bars indicate the predicted statistical error 
(half error bar corresponds to a single standard deviation).
\label{F3}
}
\end{figure}

To demonstrate this, we have performed computer simulations
of measurements and reconstructed the diagonal density-matrix
elements from a set of $5\times 10^{3}$ measured data,
on assuming the system is initially prepared in a state
$\langle n | \psi \rangle$ $\!\propto $ $\! \alpha^{n} (n!)^{-1/2}$.
The exact density-matrix elements $\varrho_{nn}$ and the exact time-averaged
position distribution are shown in Figs.~\ref{F3}(a) and \ref{F3}(b),
respectively. The position distribution reveals
a structure of two peaks located at the turning points.
The peak on the side of weaker potential is broader
than the peak in the region of strong repulsive force.
Comparing Figs.~\ref{F3}(c) and \ref{F3}(d), we see that
the reconstruction based on least-squares inversion [Fig.~\ref{F3}(c)]
yields much less fluctuating results (especially for larger values of
the quantum number $n$) than that using IWM
[Fig.~\ref{F3}(d)]. The reason is that in the relevant $x$ interval,
in which the position distribution is essentially nonzero,
the least-squares sampling functions become weakly oscillating around
zero when the quantum number $n$ is increased, whereas
larger values are attained at such $x$ values for which the position
probability is small [cf. Figs.~\ref{F2} and \ref{F3}(b)]. 
Hence, the least-squares method is suited
for extracting the information about the density-matrix elements
from the position distribution even when the position distribution
is not measured exactly. This is not the case when the
IWM is used, since the associated pattern functions
rapidly oscillate with large amplitudes over the whole
region of probable positions, so that the position distribution
must be measured with high precision in order to extract from it
the relevant information on the density-matrix elements.

%%%%%%%%%%%%%%%%%%%%%%%%%%%%%%%%%%%%%%%%%%%%%%%%%%%%%%%%%%%%%%%%%%

\section{Observational time}
\label{S3}

Let us now turn to the problem of measurement time.
In the case of an undamped harmonic oscillator the situation is
simple. Since the motion is periodic, observation of
the position distribution over one period $T$ must yield all
information about the quantum state. Moreover, taking into account the
symmetry $p(x,t+T/2)$ $\! =$ $\! p(-x,t)$, we see that
observation over one half period is already sufficient for
the quantum-state reconstruction, so that $T/2$ can be chosen
as observational time. In general the excitation energies are
not equidistant, and they are not discrete even for a bound system.
Anharmonicities prevent the energy levels from being equidistant and
dampings that accompany any motion give rise to line broadenings.

In order to approximately apply IWM to the
density-matrix reconstruction of an anharmonic bound system, it has
been suggested to choose the observational time $T$ in Eq.~(\ref{t2})
such that it is long compared with all inverse transition frequencies
\cite{LeoRaymer}. For a Morse oscillator observational times that
correspond to fractional revivals have been proposed \cite{Leo-mors}.
To be more specific, the first fractional revival for a Morse oscillator
appears for $T_{1}$ $\! =$ $\!2\pi (n_{M}+1/2)/\Omega$, where
$\Omega$ $\!=$ $\!(1-a^2/2)$.
Since the proposed observational times can be comparable or even longer
than the characteristic damping times, terms oscillating at discrete
frequencies could not be singled out in this way. The
question raises as to whether or not it is possible to reconstruct
the density matrix from data measured during a substantially
shorter time interval.

%%%%%%%%%%%%%%%%%%%%%%%%%%%%%%%%%%%%%%%%%%%%%%%%%%%%%%%%%%%%%%%%%%%%

\subsection{Factorable sampling functions}

To answer the question, we recall that owing to the finite
value of $n_{\rm max}$ there is only a finite number of 
exponential functions
\begin{eqnarray}
\label{i00}
g_{k}(t) = e^{-i\omega_{k}t} \qquad
(\omega_{k} = \omega_{n}\!-\!\omega_{n'})
\end{eqnarray}
that -- as long as
dampings can be neglected -- govern the time evolution of $p(x,t)$
[cf. Eqs.~(\ref{t1}) and (\ref{t1a})].
We can then construct sets of functions that are biorthonormal to
these exponentials in a chosen time interval in order to
appropriately decompose $p(x,t)$ and extract from the decomposition
the density-matrix elements. In principle, any interval 
(small compared with the damping times) can be used. Moreover, there
are various possibilities of constructing a biorthonormal system to a
finite set of linearly independent functions on a given interval.
A possible way is to construct them as linear combinations of the
exponentials $g_{k}(t)$, where the expansion coefficients can be
calculated using  least-squares inversion. From App.~\ref{Ap1}
[see Eqs.(\ref{e1}) and (\ref{e6a})] in a very similar way
as in the preceding section the orthonormal set of functions can
be given by
\begin{eqnarray}
\label{i01}
f_{k}(t) = \sum_{l} F_{k,l} \, g_{l}^{\ast}(t) ,
\end{eqnarray}
where ${\bf F} = {\bf G}^{-1}$ and the matrix ${\bf G}$ reads as
\begin{eqnarray}
\label{i02}
G_{k,l} = \int_{0}^{T} {d}t \, g_{k}^{\ast}(t) g_{l}(t),
\end{eqnarray}
with $T$ being the length of the chosen time interval. From the
construction of the functions $f_{k}(t)$ it is clear that
\begin{eqnarray}
\label{i03}
\int_{0}^{T} {d}t \, f_{k}(t) g_{k'}(t) = \delta_{k,k'}.
\end{eqnarray}
In place of Eq.~(\ref{t3ab}) we have
\begin{eqnarray}
\label{i03a}
p^{(k)}(x) = \int_{0}^{T} dt \ f_{k}(t) p(x,t),
\end{eqnarray}
and the reconstructed density matrix is then
\begin{eqnarray}
\label{i03b}
\tilde \varrho_{n,n'} = \int {d}x \int_{0}^{T} {d}t \,
K_{n,n'}(x)f_{k}(t) \, p(x,t),
\end{eqnarray}
with $K_{n,n'}(x)$ from Eq.~(\ref{tad3}).

%%%%%%%%%%%%%%%%%%%%%%%%%%%%%%%%%%%%%%%%%%%%%%%%%%%%%%%%%%%%%%%%%%%%%%%

\subsection{Nonfactorable sampling functions}

The matrix ${\bf G}$ in Eq.~(\ref{i02}) depends on the chosen time interval.
When the time interval is too short, then the functions $g_{k}(t)$ tend
to linearly dependent functions. The functions $f_{k}(t)$ become strongly
oscillating with large amplitudes, so that even small experimental
inaccuracies can give rise to big errors.
To get reasonable values of the statistical fluctuations, the 
required interval of time integration may be too large.
The observational time however can be
drastically reduced when the inversion of Eq.~(\ref{t1})
is performed in time and space simultaneously. Direct application
of least-squares inversion to Eq.~(\ref{t1}) yields
\begin{eqnarray}
\label{i1}
\tilde \varrho _{n,n'} = \int \! dx \int_{0}^{T}\! {d}t \,
K_{n,n'}(x,t) \, p(x,t) ,
\end{eqnarray}
where the  time- and position-dependent (nonfactorable) sampling
functions are given by
\begin{eqnarray}
\label{i2}
K_{n,n'}(x,t)
= \sum_{m,m' \! \le \! n_{\rm max}} F_{n,n';m,m'}
S_{m,m'}^{\ast}(x,t)
\end{eqnarray}
and ${\bf F}$ $\!=$ $\!{\bf G}^{-1}$,
with the matrix ${\bf G}$ being now defined by
\begin{eqnarray}
\label{i3}
G_{m,m';n,n'}
=\int \! {d}x  \int_{0}^{T} \! {d}t \,
S_{m,m'}^{\ast}(x,t)  S_{n,n'}(x,t),
\end{eqnarray}
where the functions $S_{n,n'}(x,t)$ are given by Eq.~(\ref{t1a}).

\begin{figure}[htb]
\epsfysize=23cm
\epsfbox{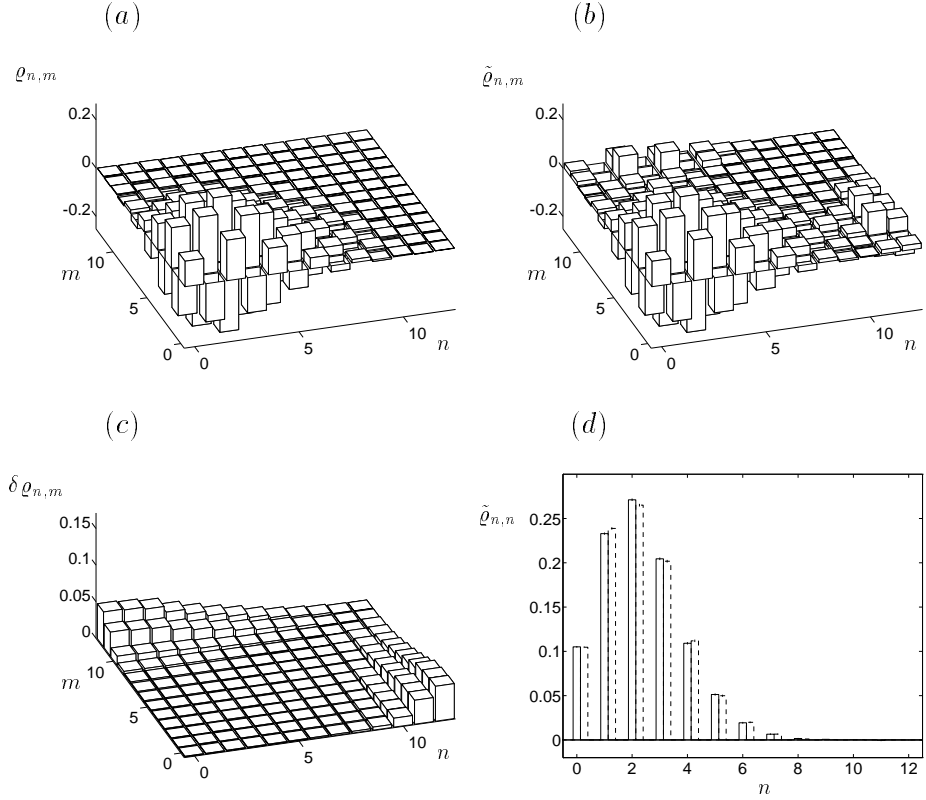}
\vspace{-2cm}
\caption{
Reconstruction of the density-matrix elements
of the same system as in Fig.~\protect\ref{F3}
from the data of a simulated position measurement, in which
at each of $N_{\rm t}$ $\!=$ $\!120$ equidistant times
$N_{\rm e}$ $\!=$ $\!5\times 10^{3}$ events are recorded, the
overall time interval being  $T$ $\!=$ $\!6 \pi/(\omega_{1}-\omega_{0})$;
exact density-matrix elements $\varrho_{n,m}$ (a),
reconstructed (real parts of the) density-matrix elements
$\tilde\varrho_{n,m}$ (b), predicted  statistical error
$\delta\varrho_{n,m}$ (c),
comparison of the diagonal elements obtained from the data recorded
during the actual measurement time $T$ (full lines) with the
diagonal elements obtained (for the same total number
of events) in the case when $T$ $\!\to$ $\!\infty$ (dashed lines) (d).
\label{F4}
}
\end{figure}

Results of reconstruction of the density matrix within
the framework of Eqs.~(\ref{i1}) -- (\ref{i3}) are plotted in
Fig.~\ref{F4}
for the same system as in Fig.~\ref{F3}.
In our computer experiments
we have assumed that measurements at $N_{\rm t}$ $\!=$ $\!120$
times in a (relatively small) time interval $T$ $\!=$
$\!6\pi /(\omega_{1}\!-\!\omega_{0})$ are performed and
$N_{\rm e}$ $\!=$ $\!5\times 10^{3}$ events at each time are recorded.
The grid of measurement points is chosen such that
the systematical error owing to discretization is reduced
below the statistical one. It should be noted that for chosen value
of $N_{\rm t}$ both the systematical and the statistical errors of the
off-diagonal density-matrix elements increase with the ``distance'' from
the diagonal elements [for the statistical error, see
Fig.~\ref{F4}(c)]. From Figs.~\ref{F4}(a) -- \ref{F4}(d) we see
that least-squares reconstruction yields a good estimation of the
density-matrix elements even for measurement times much shorter than
the (first) fractional-revival time $\approx 24 \pi
/(\omega_{1}-\omega_{0})$.
In particular, Fig~\ref{F4}(d)
reveals that the accuracy of the reconstruction is comparable
with the accuracy that can be achieved -- for a comparable number of
total events -- in the case when the observational time is extended
to infinity.

It should be mentioned that the reconstruction scheme may advantageously
be applied also to harmonic oscillators. An
example may be balanced homodyne detection of radiation-field modes.
Here the phase interval in which the quadrature-component distributions
are measured can be reduced below a $\pi$ interval. Recently it was
suggested to transfer the formalism of density-matrix reconstruction 
of quantum harmonic oscillators to
classical tomographical reconstruction of objects with space-varying
transparencies \cite{Ariano-OC}.
This can be advantageous when the probing beam is very weak and
a relatively small number of data is available. Application of the
least-squares inversion with shorter intervals of the angular variable
can make the method also suitable for tomographical reconstruction of
objects whose projections on some directions are not available.

It is worth noting that the assumption that the measurements are
performed over the whole $x$ axis may also be dropped. When the
available $x$ interval is limited, then -- in close analogy to limited
time intervals -- the $x$ integrals in the reconstruction formulas can be
reduced to this interval. Hence, the least-squares inversion method
enables one to reconstruct the density matrix also in cases when the time
and ``position'' intervals are smaller than the theoretically allowed ones.

%%%%%%%%%%%%%%%%%%%%%%%%%%%%%%%%%%%%%%%%%%%%%%%%%%%%%%%%%%%%%%%%%%%%%%%

\subsection{Inclusion of damping}

So far we have assumed that damping can be disregarded on
the chosen time scale. However, the method can also be
extended to damped systems. In this case the dependence on time
of the coefficient functions $S_{n,n'}(x,t)$ in Eq.~(\ref{t1a})
is given by appropriately decaying functions in place of
purely oscillating exponentials $\exp[-i(\omega_{n}\!-\!\omega_{n'})t]$.
Suppose that the density matrix evolves according to some master equation
\begin{eqnarray}
\label{iG0}
\dot{\!\hat\varrho} = \hat{\cal L} \hat\varrho,
\end{eqnarray}
where $\hat{\cal L}$ is a linear superoperator,
\begin{eqnarray}
\label{iG0a}
\hat{\cal L} \hat\varrho = \frac{1}{i\hbar}[\hat H,\hat\varrho]
+ \hat {\cal R} \hat\varrho,
\end{eqnarray}
with $\hat{\cal R}$ describing the effect of (Markovian) damping.
Since the solution of a master equation (\ref{iG0}) can always be
represented in the form of
\begin{eqnarray}
\label{iG1}
\varrho_{m,m'}(t) = \sum_{n,n'} U_{m,m';n,n'}(t) \, \varrho_{n,n'},
\end{eqnarray}
$\varrho_{n,n'}$ $\!\equiv$ $\!\varrho_{n,n'}(0)$,
the above given reconstruction formulas, such as Eqs.~(\ref{i1}) --
(\ref{i3}) remain valid when the functions $S_{n,n'}(x,t)$ given
in Eq.~(\ref{t1a}) are replaced with
\begin{eqnarray}
\label{iG2}
S_{n,n'}(x,t) = \sum_{m,m'} \psi_{m}(x) \psi_{m'}(x) U_{m,m';n,n'}(t).
\end{eqnarray}

It should be pointed out that in some cases it may happen that
the ${\bf G}$ matrix is quasi-singular so that it cannot be inverted
in the usual way. Physically it means that the available data do not
carry enough information about some of the density-matrix elements.
In this case regularized inversion can be used, as discussed in
App.~\ref{Ap1} [see Eqs.~(\ref{e8}) and (\ref{e9})].

%%%%%%%%%%%%%%%%%%%%%%%%%%%%%%%%%%%%%%%%%%%%%%%%%%%%%%%%%%%%%%%%%%

\section{Data smearing}
\label{S4}

In general, an $x$ measurement can only be performed with
finite accuracy and the time cannot be controlled precisely,
so that in practice one always deals with more or less smeared data.
For example, in optical homodyne tomography nonperfect
detection yields smeared quadrature-component distributions. 
In molecular emission tomography \cite{Dunn2,Kowalczyk}
time-resolved fluorescence spectra are measured, which 
necessarily implies that a perfect resolution of frequency 
and time is impossible.
Whereas attempts have been made to compensate for nonperfect detection
in optical homodyne tomography \cite{Ariano,Herzog}, an open problem
has been the inclusion of data smearing in quantum-state reconstruction
for more general systems.

Mathematically, smearing means that instead of $p(x,t)$ a convolution
\begin{eqnarray}
\label{r1}
\bar p(x,t) = \int \!{d}x' \int \! {d}t' \,  V(t'-t)
W(x'-x) p(x',t')
\end{eqnarray}
is typically measured, where $V(t)$ and $W(x)$ are single-peak functions
(centered at zero) describing the time and position windows.
Recalling Eqs.~(\ref{t1}) and (\ref{t1a}), $\bar p(x,t)$ can
be related to the density-matrix elements as
\begin{eqnarray}
\label{r2}
\bar p(x,t) = \sum_{n,n' \le n_{\rm max}}
\bar S_{n,n'}(x,t)
\, \varrho _{n,n'} ,
\end{eqnarray}
where
\begin{eqnarray}
\label{r2a}
\bar S_{n,n'}(x,t)  =  V_{n,n'}(t) W_{n,n'}(x),
\end{eqnarray}
with
\begin{eqnarray}
\label{r4}
V_{n,n'}(t) = \int {d}t' \, e^{-i(\omega _{n} - \omega _{n'})t'} V(t'-t)
\end{eqnarray}
and
\begin{eqnarray}
\label{r3}
W_{n,n'}(x) = \int {d}x' \, \psi_{n}(x') \psi_{n'}(x') W(x'-x).
\end{eqnarray}
The inversion of Eq.~(\ref{r2}) can be done in the same way
leading to Eqs.~(\ref{i1})-(\ref{i3}), i.e.,
\begin{eqnarray}
\label{r5}
\tilde \varrho _{n,n'} = \int {d}x \int {d}t \, \bar K_{n,n'}(x,t) \,
\bar p(x,t) ,
\end{eqnarray}
where $\bar K_{n,n'}(x,t)$ is calculated according to Eq.~(\ref{i2})
[together with Eq.~(\ref{i3})], but with $\bar S_{n,n'}(x,t)$ in place
of $S_{n,n'}(x,t)$. The limits of integration in
Eq.~(\ref{r5}) are given by the chosen intervals of measurement.
In practice $\bar p(x,t)$ is usually measured on a grid of points
$\{x_{l},t_{k}\}$, so that the integrals over $x$ and $t$ in
Eq.~(\ref{r5}) (and the integrals determining the $\bf G$ matrix)
are replaced with sums.

It should be noted that due to data smearing it may be necessary to 
perform a regularized inversion as mentioned above. For example, when
the measured data are relatively insensitive to temporal changes
of the quantum state, then it may be better to set some reconstructed
density-matrix elements close to zero rather than to let them
become very large, ``trying to fit the data'' as close as possible.
Clearly, one must correctly interpret the result -- instead of claiming
that the elements are measured to be zero one should say that there
is not enough evidence for nonzero values.

\begin{figure}[htb]
\epsfysize=23cm
\epsfbox{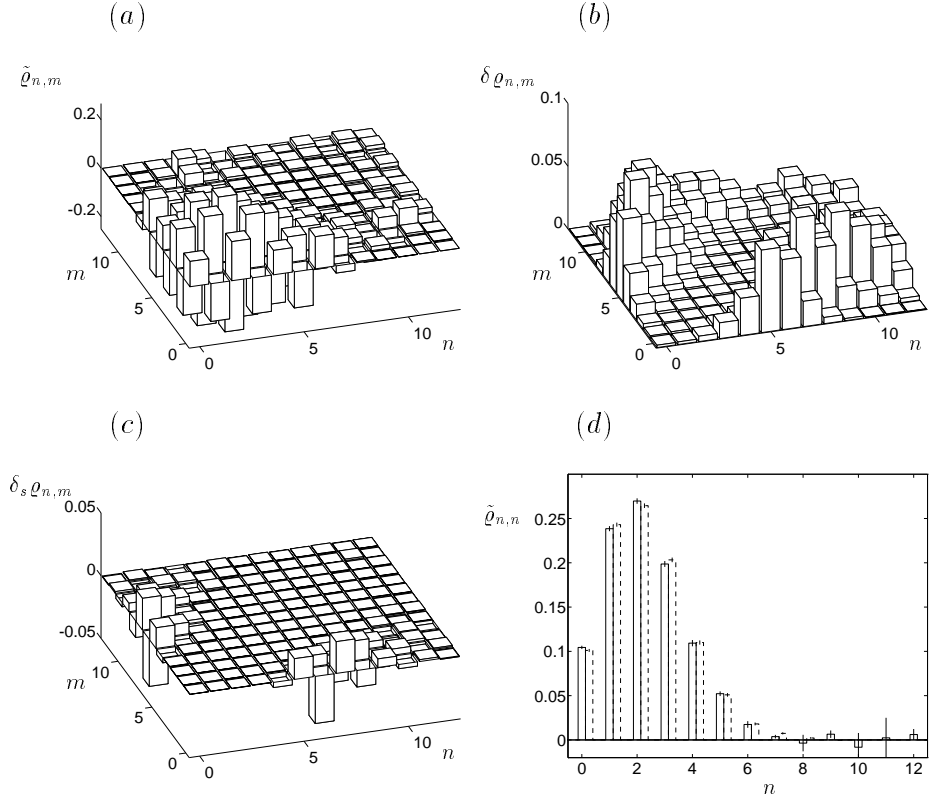}
\vspace{-3cm}
\caption{
Reconstruction of the density-matrix elements
of the same system as in Fig.~\protect\ref{F3}
from a simulated measurement of the smeared distribution $\bar p(x,t)$
[$\sigma _{t}$ $\!=0.2$ $\!\pi /(\omega_{1}-\omega_{0})$,
$\sigma_{x}$ $\!=$ $\!0.3$] at $N_{\rm t}$ $\!=$ $\!30$ equidistant times
[during the observational time $T$ $\!=$ $\!6 \pi /(\omega_{1}-\omega_{0})$]
and $N_{\rm x}$ $\!=$ $\!15$ equidistant positions
[in the interval $-2$ $\!\leq$ $\!x$ $\!\leq$ $\!10$] for a total number
of events of $N_{\rm tot}$ $\!=$ $\!10^{5}$, using Tikhonov
regularization with $\lambda$ $\! =$ $\! 2 \times 10^{-3}$;
reconstructed (real parts of the) density-matrix elements
$\tilde\varrho_{n,m}$ (a),
predicted  statistical error $\delta\varrho_{n,m}$ (b),
systematical error $\delta_{\rm s}\varrho_{n,m}$ due to regularization (c),
comparison of the diagonal elements obtained from the smeared data
recorded during the actual measurement time $T$ (full lines) with the
diagonal elements obtained (for the same total number of events)
without data smearing, and $T$ $\!\to$ $\!\infty$ (dashed lines) (d).
\label{F7}
}
\end{figure}

\begin{figure}[htb]
\epsfysize=23cm
\epsfbox{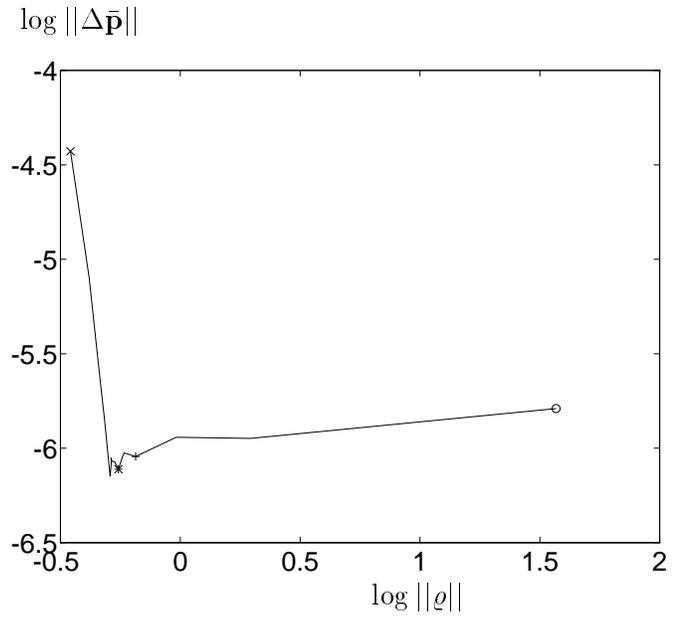}
\vspace{-2cm}
\caption{
The $L$ curve for the problem considered in Fig.~\protect\ref{F7};
$\lambda = 10^{-4}$ ($o$),
$2\times 10^{-3}$ ($+$),
$5\times 10^{-3}$ ($*$),
$5 \times 10^{-2}$ ($\times$).
\label{F9}
}
\end{figure}

The application of the method to systems with data smearing is
illustrated in Fig.~\ref{F7} for the same system
as in Fig.~\ref{F4}. In the computer simulations of measurements
the data are assumed to be smeared over Gaussian windows,
$V(t)$ $\! =$ $\! \exp [-t^{2}/(2\sigma_{t}^{2})]$ and
$W(x)$ $\! =$ $\! \exp [-x^{2}/(2\sigma_{x}^{2})]$,
with $\sigma _{t}$ $\!=$ $\!0.2 \pi /(\omega_{1}-\omega_{0})$
and $\sigma_{x}$ $\!=$ $\!0.3$. It is further assumed that during an
observational time $T$ $\!=$ $\!6 \pi /(\omega_{1}-\omega_{0})$
measurements are performed at $N_{\rm t}$ $\!=$ $\!30$ (equidistant) times
and $N_{\rm x}$ $\!=$ $\!15$ (equidistant) positions in an
interval $-2$ $\!\leq$ $\!x$ $\!\leq$ $\!10$, the total number of
recorded events being $N_{\rm tot}$ $\!=$ $\!10^{5}$.
Hence, at a given point $(x_{l},t_{k})$ of the chosen
grid the number of recorded events is approximately given by
\begin{eqnarray}
\label{r8}
n_{l,k} = \frac{N_{\rm tot}}{T} \int \! dx \int \! dt \
W(x-x_{l}) V(t-t_{k}) p(x,t).
\end{eqnarray}
Since owing to time smearing fast oscillating terms can hardly be
resolved, the measured distribution $\bar p(x,t)$, Eq.~(\ref{r2}),
becomes insensitive to off-diagonal density-matrix elements that are
far from the diagonal. To obtain reasonable results, regularized inversion
of Eq.~(\ref{r2}) is necessary. Using the Tikhonov regularization
(see App.~\ref{Ap1}), the regularization parameter used in Fig.~\ref{F7} is
$\lambda$ $\! =$ $\! 2 \times 10^{-3}$. It should be noted that with
increasing value of $\lambda$ the statistical error [Fig.~\ref{F7}(b)]
decreases, but the systematical error [Fig.~\ref{F7}(c)] increases.
Suitable values of $\lambda$ can be obtained from the $L$ curve 
(see App.~\ref{Ap1}) of the
problem which is plotted in Fig.~\ref{F9}. The best values of $\lambda$
correspond to points near the corner.
Using singular-value decomposition in place of Tikhonov regularization,
similar results can be obtained.

A comparison with results based on measurements on the whole
time and position scales and without data smearing is shown
in Fig.\ref{F7}(d) for the diagonal density-matrix elements.
We see that for smaller values of $n$ the results are almost equally good,
however with increasing $n$ the statistical error of the density-matrix
elements reconstructed from the smeared data strongly increases.
This can be understood from the fact that typical features
(such as rapid oscillations and larger values of $x$)
of higher-excited state contributions
to $p(x,t)$ are less available from the smeared data confined to a
limited interval.

\begin{figure}[htb]
\epsfysize=23cm
\epsfbox{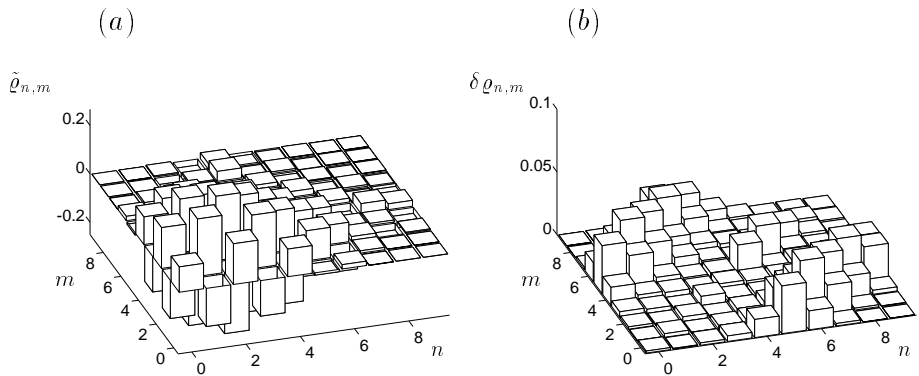}
\vspace{-2cm}
\caption{
Reconstruction of the density-matrix elements
of a system of the type considered in Fig.~\protect\ref{F3}, but
with smaller anharmonicity, $a$ $\!=$ $\!0.15$, and a truncation parameter
smaller than the number of bound states, $n_{\rm max}$ $\!=$ $\!9$,
from a simulated measurement of the smeared distribution $\bar p(x,t)$
[$\sigma _{t}$ $\!=0.2$ $\!\pi /(\omega_{1}-\omega_{0})$,
$\sigma_{x}$ $\!=$ $\!0.3$] at $N_{\rm t}$ $\!=$ $\!30$ equidistant times
[during the observational time $T$ $\!=$ $\!4 \pi /(\omega_{1}-\omega_{0})$]
and $N_{\rm x}$ $\!=$ $\!15$ equidistant positions
[in the interval $-2$ $\!\leq$ $\!x$ $\!\leq$ $\!10$] for a total number
of events of $N_{\rm tot}$ $\!=$ $\!10^{6}$, using Tikhonov
regularization with $\lambda$ $\! =$ $\!10^{-3}$;
reconstructed (real parts of the) density-matrix elements
$\tilde\varrho_{n,m}$ (a),
predicted  statistical error $\delta\varrho_{n,m}$ (b).
\label{F10}
}
\end{figure}

In the examples given above we have included in the
reconstruction of the density matrix all bound states of the
anharmonic oscillator ($n_{\rm max}$ $\!=$ $\!n_{M}$).
If there is some {\em a priori} knowledge on the quantum state
of the system, it may be possible to choose $n_{\rm max}$ such that
$n_{\rm max}$ $\!<$ $\!n_{M}$. In this way the dimension
of the matrix that must be inverted can be reduced. An advantage
is that the statistical error can also be reduced.
To illustrate such a case, in Fig.~\ref{F10} we have applied the
reconstruction scheme to a Morse oscillator whose anharmonicity
($a$ $\!=$ $\!0.15$) is smaller than that in Figs. \ref{F3} -- \ref{F7},
so that $n_{M}$ $\!=$ $\!43$.
Assuming that the system is again prepared in a state
of the form considered in Figs. \ref{F3} -- \ref{F7}, with the same
parameter $\alpha$ $\!=$ $\!-1.5$, we have set $n_{\rm max}$ $\! =$ $\! 9$.
With increasing $n_{\rm max}$ the systematical error (owing to
truncation) can be decreased, but the statistical error is increased.
Therefore, for given number of available data (and given state) one expects
that there is an optimum value of $n_{\rm max}$ for which the systematical
error is just below the statistical one.

%%%%%%%%%%%%%%%%%%%%%%%%%%%%%%%%%%%%%%%%%%%%%%%%%%%%%%%%%%%%%%%%%%

\section{Summary and conclusions}
\label{S5}

We have considered the problem of reconstruction of the density
matrix $\varrho_{n,m}$ from the data available in realistic experiments.
The studied systems can be linear oscillators as well as more general
systems with non equidistant energy spectra.
To obtain the complete quantum state, the measured quantities must be
related to all (nonvanishing) density-matrix elements that must
necessarily be known for characterizing the state. Then the problem
to be solved is the inversion of sets of linear equations.
Having enough experimental data, such a system of linear equations
can be overdetermined with respect to the density-matrix elements
sought. This enables one to perform the inversion in different ways.
Here we have applied the least-squares method. The density-matrix
elements are obtained as linear combinations of the observed
quantities, so that they fit the measured data as close as possible.
The main features of the method can be summarized as follows.

(i) The application of the method is not restricted to
the reconstruction of the density matrix from certain specific
quantities, such as the time-dependent position distribution considered
in IWM. Least-squares inversion can always be applied when
there are measurable quantities that can be related to all density-matrix
elements that significantly contribute to the quantum state of the system.
Moreover, it is not necessary that the system evolves undamped,
as it is required for applying IWM.

(ii) The flexibility of the method enables one to take into account
typical experimental effects and necessities, such as data smearing
and restrictions that limit the size of measurement intervals
and the amount of available data. For example, using
least-squares inversion and reconstructing the
density matrix of a particle moving in a Morse potential
from the time-dependent position distribution, the measurement time can
be substantially shorter than in IWM method.

(iii) Both the reconstruction of the density matrix and the estimation
of the statistical error can be performed in real time.
It is worth noting that least-squares inversion can advantageously be used
in order to reduce the statistical error below the level given by
IWM for a comparable amount of data.

(v) If the measured data are insensitive to certain density-matrix elements,
so-called regularized solutions can be constructed. That is to say,
the reconstructed values of those density-matrix elements are set close
to zero instead of dealing with strongly fluctuating values.
Regularization decreases the statistical error
of the reconstructed density-matrix elements,
but simultaneously causes a systematical error.
It is therefore necessary to
optimize the regularization such that the introduced systematical error is
below the statistical one.

(vi)
The method is essentially based on the fact that
for any physical quantum state the density-matrix elements
$\varrho_{n,m}$ must eventually decrease indefinitely with
increasing $n(m)$, so that the density matrix can effectively be
truncated at some large (but finite) value of $n(m)$.
Clearly, truncation always introduces a systematical error.
In practice it is sufficient to choose a truncation parameter
for which -- similar to regularization  -- the systematical
error is smaller than the statistical one.

Finally, it should be mentioned that there have been
other approaches to the problem of reconstruction of
the quantum state from finite sets of measured quantities and
recorded data \cite{Buzek,Wiedemann,Hradil}. In particular, the
entropic reconstruction scheme studied in \cite{Buzek,Wiedemann}
may be used to systematically reconstruct the density matrix,
starting from only a few number of quantities and data and extending
the scheme to larger sets.
In the method suggested in \cite{Hradil} conditions are imposed
on the reconstruction procedure such that the reconstructed object
is really a density matrix (i.e., the diagonal elements must not be
negative and the trace is equal to unity). Note that the
least-squares method -- and other methods that are based
on linear transforms of measured data -- can produce ``negative
probabilities'' resulting from experimental inaccuracies.
The reconstruction schemes in \cite{Buzek,Wiedemann,Hradil}
are based on the solution of sets of nonlinear equations
whose solution may become extremely difficult when large sets of data
must be processed.
In contrast, the least-squares method enables one
to reconstruct the density-matrix elements in real time
in a very straightforward way. Such excesses as ``negative probabilities''
can be kept within the statistical error bars whose magnitude can
easily be estimated and eventually decreased by increasing the
number of measurements.

\section*{Acknowledgments}

This work was supported by the Deutsche Forschungsgemeinschaft.
One of us (T.O.) is grateful to U. Leonhardt for discussion
about the Morse-oscillator problem and to S. Wallentowitz for
advises to least-squares inversion.

%%%%%%%%%%%%%%%%%%%%%%%%%%%%%%%%%%%%%%%%%%%%%%%%%%%%%%%%%%%%%%%%%%%%%

\appendix
\section{Elements of least-squares inversion}
\label{Ap1}

To give a summary of least-squares inversion \cite{Robin,Tan},
let us consider a (possibly unknown) $n_{0}$ dimensional ``state''
vector ${\bf f}$ and a stochastic linear transform
\begin{eqnarray}
\label{e1}
{\bf y} = {\bf A \ f} + {\bf n}
\end{eqnarray}
yielding an $m_{0}$ dimensional ($m_{0}$ $\! \ge $ $\! n_{0}$) ``data''
vector $\bf y$ available from measurements. Here ${\bf A}$ is a given
$m_{0}$ $\! \times$ $\! n_{0}$ matrix, and
${\bf n}$ is an $m_{0}$ dimensional vector whose random elements
with zero means and covariance matrix ${\bf W}^{-1}$
describe the noise associated with realistic measurements.
The task is to infer ${\bf f}$ from ${\bf y}$.

  From the point of view of Bayesian inference, probability
distributions for ${\bf y}$ and ${\bf f}$ can be introduced,
and the probability for a state vector ${\bf f}$ under the condition
that there is a date vector ${\bf y}$ can be given by
\begin{eqnarray}
\label{e2}
P({\bf f}|{\bf y}) \propto P({\bf y}|{\bf f}) P({\bf f}),
\end{eqnarray}
where $P({\bf y}|{\bf f})$ is the corresponding conditional probability
for the date vector ${\bf y}$,
and $P({\bf f})$ is an {\em a priori} probability for the
state vector ${\bf f}$. We then look for the state vector ${\bf f}$
for which $P({\bf f}|{\bf y})$ is maximum.
Assuming that the noise is Gaussian, we have
\begin{eqnarray}
\label{e3}
P({\bf y}|{\bf f}) \propto \exp\!\left[ -{\textstyle\frac{1}{2}}
({\bf y}-{\bf Af})^{\dag} {\bf W} ({\bf y}- {\bf Af})
\right] .
\end{eqnarray}
The {\em a priori} probability $P({\bf f})$ represents our
knowledge of the state
when no data are available. Hence $P({\bf f})$
can be set constant if no
state
is to be preferred.
Under the conditions made maximization of $P({\bf f}|{\bf y})$
is equivalent to minimization of
\begin{eqnarray}
\label{e4}
C({\bf f}) = ({\bf y}-
{\bf Af})^{\dag} {\bf W} ({\bf y}- {\bf Af}),
\end{eqnarray}
from which we see that the minimum is attained at
${\bf f}$ $\!=$ $\! \tilde {\bf f}$ satisfying the equation
\begin{eqnarray}
\label{e5}
{\bf A}^{\dag}{\bf W}{\bf A}\tilde {\bf f}
={\bf A}^{\dag}{\bf W}{\bf y}.
\end{eqnarray}
If ${\bf W}$ is diagonal (i.e., the noise is uncorrelated), then
Eq.~(\ref{e4}) represents a sum of weighted squares of the differences
between the components of the data vector ${\bf y}$ and the components of
the transformed vector ${\bf A f}$, each term of the sum being
multiplied by a weight given by the corresponding element of ${\bf W}$.
Provided that ${\bf A}^{\dag}{\bf W}{\bf A}$ is not singular, from
Eq.~(\ref{e5}) we obtain
\begin{eqnarray}
\label{e6}
\tilde {\bf f}=({\bf A}^{\dag}{\bf W}{\bf A})^{-1}
{\bf A}^{\dag}{\bf W}{\bf y}.
\end{eqnarray}
Otherwise, the inversion of
Eq.~(\ref{e5}) is not unique and further criteria must be used to
select a solution.
When the matrix ${\bf W}$ is not known, then
it may be set a multiple of a unity matrix, so that
Eq.~(\ref{e6}) reduces to
\begin{eqnarray}
\label{e6a}
\tilde {\bf f}=({\bf A}^{\dag}{\bf A})^{-1}
{\bf A}^{\dag}{\bf y},
\end{eqnarray}
provided that ${\bf A}^{\dag}{\bf A}$ is not singular. Equation
(\ref{e6a}) still gives the correct averaged inversion, but the
statistical fluctuation of the result may be (slightly) enhanced.

If the data are not sensitive enough to some state-vector components,
then these components can hardly be determined with reasonable
accuracy. Mathematically, ${\bf A}^{\dag}{\bf W}{\bf A}$ becomes
(quasi-)singular and regularizations, such as Tikhonov
regularization and singular-value decomposition, are required
to solve approximately Eq.~(\ref{e5}). For simplicity let us set
${\bf W}$ $\!=$ $\!{\bf I}$, where ${\bf I}$ is the unity matrix.

Using Tikhonov regularization, it is assumed that in the absence
of significant data some components of the state vector can be preferred
by a properly chosen {\em a priori} probability $P({\bf f})$.
Assuming a Gaussian distribution and preferring the components close to
zero (if information about them is lacking), we may write
\begin{eqnarray}
\label{e7}
P({\bf f}) \propto \exp\!\left( - {\textstyle\frac{1}{2}} \lambda ^2
{\bf f}^{\dag} {\bf f}\right),
\end{eqnarray}
where the (positive) parameter $\lambda$ is a measure of the
strength of regularization. Maximization of $P({\bf f}|{\bf y})$
then yields, on recalling Eqs.~(\ref{e2}) and (\ref{e3}),
\begin{eqnarray}
\label{e8}
\tilde {\bf f}=(\lambda ^{2} {\bf I} + {\bf A}^{\dag}{\bf A})^{-1}
{\bf A}^{\dag}{\bf y}.
\end{eqnarray}
Note that $(\lambda ^{2} {\bf I} + {\bf A}^{\dag}{\bf A})$
has only positive eigenvalues and is thus always invertible.
A possible choice of
$\lambda$
is based on the so-called $L$ curve, which is a log-log plot of
$||{\bf f} ||$
versus $||\Delta {\bf y} ||$,
$\Delta {\bf y}$ $\!=$ $\!{\bf y}-{\bf A}{\bf f}$,
for different values of
$\lambda$.
The points on the horizontal branch correspond to large noise,
whereas the points on the vertical branch correspond to
large data misfit. Optimum choice of $\lambda$ corresponds to points
near the corner of the $L$ curve.

Applying singular-value decomposition, the inversion of the matrix
$({\bf A}^{\dag}{\bf A})$ is performed such that their eigenvalues
whose absolute values are smaller than the (positive) parameter 
of regularization $\sigma_{0}$ are treated as zeros, but the inversions 
are set zero (instead to infinity). This operation is called 
``pseudoinverse'' of a matrix,
\begin{eqnarray}
\label{e9}
\tilde {\bf f}= {\rm Pseudoinverse}({\bf A}^{\dag}{\bf A}; \sigma_{0})
{\bf A}^{\dag}{\bf y}.
\end{eqnarray}
For $\sigma _{0}$ close to zero the result of Eq.~(\ref{e9}) is
similar to that of Eq.~(\ref{e6a}). With increasing $\sigma _{0}$
smaller absolute values of components of $\tilde {\bf f}$ are preferred.

The effect of the regularization parameters $\lambda$ and $\sigma _{0}$
is similar. The statistical error of the reconstructed
state vector $\tilde {\bf f}$ is decreased, but simultaneously bias
towards zero is produced.
Hence, optimum parameters are those for which
the bias is just below the statistical fluctuation.
The bias can be estimated, e.g., by Monte Carlo generating new sets of
``synthetic'' data from the reconstructed state. From these sets one can
again reconstruct new sets of $\tilde {\bf f}$.
The difference between the mean value of
the states reconstructed from the synthetic data
and the originally reconstructed state estimates the bias.

%%%%%%%%%%%%%%%%%%%%%%%%%%%%%%%%%%%%%%%%%%%%%%%%%%%%%%%%%%%%%%%%%%%%%%%%%%

\section{Calculation of the statistical error}
\label{Ap2}

Let us suppose that the probabilities for observing a physical quantity,
$p_{l}(s)$, can be related to quantities $f_{k}$ as
\begin{eqnarray}
\label{e10}
p_{l}(s) = \sum_{k} A_{l,k}(s) \, f_{k},
\end{eqnarray}
where $l$ refers to the values of the quantity to be observed and
$s$ is some parameter that can be changed during the observation.
Measuring $p_{l}(s)$ for all $l$ and $s$ values, we may
identify the measured values $\tilde p_{l}(s)$ with a data vector,
from which -- according to Eqs.~(\ref{e6a}) or (\ref{e8}) or (\ref{e9}) --
the quantities $\tilde f_{k}$ can be reconstructed,
\begin{eqnarray}
\label{e12}
\tilde f_{k} = \sum_{l,s} K_{k,l}(s) \, \tilde p_{l}(s).
\end{eqnarray}
The measured probabilities can be given by
$\tilde p_{l}(s)$ $\! =$ $\! n_{l}(s)/N(s)$, where (for chosen $s$)
$N(s)$ and $n_{l}(s)$, respectively, are the total number of events
and the number of events yielding the result $l$.
Assuming that $n_{l}(s)$ has approximately a Poissonian distribution
with mean and variance equal to $N(s)p_{l}(s)$, the mean and variance of
$\tilde f_{k}$ can easily be calculated, namely,
\begin{eqnarray}
\label{e13}
\langle {\rm Re} \tilde f_{k} \rangle
& = & \sum_{l,s}{\rm Re}[ K_{k,l}(s)] \,
\frac{\langle n_{l}(s)\rangle}{N(s)}
\nonumber \\
& = & \sum_{l,s}{\rm Re}[ K_{k,l}(s)] \, p_{l}(s)
\end{eqnarray}
and
\begin{eqnarray}
\label{e14}
{\rm Var}( {\rm Re}\tilde f_{k}) & = &
\sum_{l,s}{\rm Re}[ K_{k,l}(s)]^{2} \,
\frac{{\rm Var }[n_{l}(s)]}{N^{2}(s)}
\nonumber \\
& = & \sum_{l,s} {\rm Re}[K_{k,l}(s)]^{2} \,
\frac{p_{l}(s)}{N(s)} \, ,
\end{eqnarray}
and for the imaginary part accordingly.
In practice, the unknown exact probabilities $p_{l}(s)$
are replaced with the estimates $\tilde p_{l}(s)$.

%%%%%%%%%%%%%%%%%%%%%%%%%%%%%%%%%%%%%%%%%%%%%%%%%%%%%%%%%%%%%%%%%%%

\onecolumn
\widetext

\newpage

\end{document}